\RequirePackage{caption}
\let\abovecaptionskip\relax
\let\belowcaptionskip\relax
\documentclass[conference]{IEEEtran}
\usepackage{url,hyperref, hyphenat}
\IEEEoverridecommandlockouts
\usepackage{cite}
\usepackage{amsmath,amssymb,amsfonts}
\usepackage{algorithmic}
\usepackage{textcomp}
\usepackage{xcolor}
\usepackage{mathtools}
\usepackage{tikzpagenodes}

\newtheorem{definition}{Definition}
\usepackage{graphicx,amssymb,braket,xcolor,subfig,upgreek,bbold, numprint}
\npthousandsep{,}
\def\BibTeX{{\rm B\kern-.05em{\sc i\kern-.025em b}\kern-.08em
    T\kern-.1667em\lower.7ex\hbox{E}\kern-.125emX}}
\begin{document}
\IEEEpubid{\begin{minipage}[t]{\textwidth}\ \\[10pt]
        \normalsize{978-1-6654-8027-7/22/\$31.00 \copyright 2023 IEEE}
\end{minipage}} 
\title{A Meta Path-based Approach for Rumor Detection on Social Media\\
}
\author{\IEEEauthorblockN{Bita Azarijoo}
\IEEEauthorblockA{\textit{Faculty of New Sciences}\\
\textit{and Technologies} \\
\textit{University of Tehran}\\
Tehran, Iran \\
azarijoo@ut.ac.ir}
\and
\IEEEauthorblockN{Mostafa Salehi}
\IEEEauthorblockA{\textit{Faculty of New Sciences and Technologies}\\
\textit{University of Tehran}\\
\textit{School of Computer Science, IPM Institute} \\
 \textit{ for Research in Fundamental Sciences}\\
Tehran, Iran  \\
mostafa.salehi@ut.ac.ir}
\and
\IEEEauthorblockN{Shaghayegh Najari}
\IEEEauthorblockA{\textit{Faculty of New Sciences}
\\ \textit{and Technologies} \\
\textit{University of Tehran}\\
Tehran, Iran  \\
najari.shaghayegh@ut.ac.ir}
}

\maketitle

\begin{abstract}
The prominent role of social media in people's daily lives has made them more inclined to receive news through social networks than traditional sources. This shift in public behavior has opened doors for some to diffuse fake news on social media; and subsequently cause negative economic, political, and social consequences as well as distrust among the public.

There are many proposed methods to solve the rumor detection problem, most of which do not take full advantage of the heterogeneous nature of news propagation networks. With this intention, we considered a previously proposed architecture as our baseline and performed the idea of structural feature extraction from the heterogeneous rumor propagation over its architecture using the concept of meta path-based embeddings. We named our model Meta Path-based Global Local Attention Network (MGLAN). Extensive experimental analysis on three state-of-the-art datasets has demonstrated that MGLAN outperforms other models by capturing node-level discrimination to different node types.\end{abstract}


\section{Introduction}\label{sec:introduction}
Nowadays, interactions with social networks have become an inseparable part of people's lives for their ease of use and fast dissemination of information on a global scale. In this regard, in 2012, only 45\% of people used social media to access news, whereas  this number jumped to 65\% in 2016 \cite{shu2017fake}. Also, the 2020 Covid-19 pandemic caused a 30\% growth in Twitter daily usage \cite{paka2021cross}. Unfortunately, this rapid increase in using social media has provided an opportunity for vicious users to spread fake news to cause serious individual, economic, and political repercussions \cite{zhou2020survey}. For example, in 2013,  the Associated Press (AP) account on Twitter was hacked, and a piece of news was published claiming an explosion occurred in the White House and Barak Obama was injured. 
Although the publishing account discredited this rumor within seconds, it leveraged through Twitter and caused the stock value to fall by  130 billion dollars \cite{shu2017fake}. Consequently, studying fake news and preventing its dissemination as soon as possible is yet an active and open research area. 

To avoid the detrimental effects of fake news, there are websites like Snopes\footnote{https://www.snopes.com}, GossipCop\footnote{https://www.gossipcop.com}, and Poitifact\footnote{https://www.politifact.com}, but they can not detect fake news automatically in their early stages of propagation and rely on manual user intervention for fact-checking as well. As a consequence, the detection of fake news can be  a time-consuming process. Thus, to solve this issue under a real scenario, various machine learning-based methods have been proposed, many of which depend on analyzing text to extract language styles of fake news, but text in social media has a short length, and we face the data sparsity problem. Other methods like  CSI \cite{ruchansky2017csi} need many user responses, which is time-consuming. It also models the propagation path of retweets as a sequence. Other models like GLAN \cite{yuan2019jointly} model both local and global relations of news and users but fails to capture intrinsic structural difference among node types when we do not have access to manually-extracted node (user, tweet, etc.) features. Our research question underlies the fact that if the metadata of users and tweets are unavailable, we would be able to extract structural features considering the difference in node types. Therefore, The contributions of our work are:
\begin{itemize}
    \item We select GLAN as our baseline model and aim to capture meaningful structural embeddings using the concepts of meta paths in heterogeneous news propagation networks.
    \item Our experiments on three real-world datasets demonstrate improvements over previous works in terms of accuracy and F1 score.
\end{itemize}

For the rest of the paper, in section \ref{sec:related_work}, we overview related works for solving rumor detection problem. In section \ref{sec:preliminaries}, we define preliminaries and formulate our problem. In section \ref{sec:proposed_method}, we describe all components of the baseline paper as well as our idea to improve upon it. In section \ref{sec:evaluation}, we evaluate our method on three real-world datasets. Finally, in section \ref{sec:conclusion},  all future works applicable to our work are proposed.

\begin{figure*}[ht]
  \begin{center}
    \includegraphics[width=0.75\textwidth]{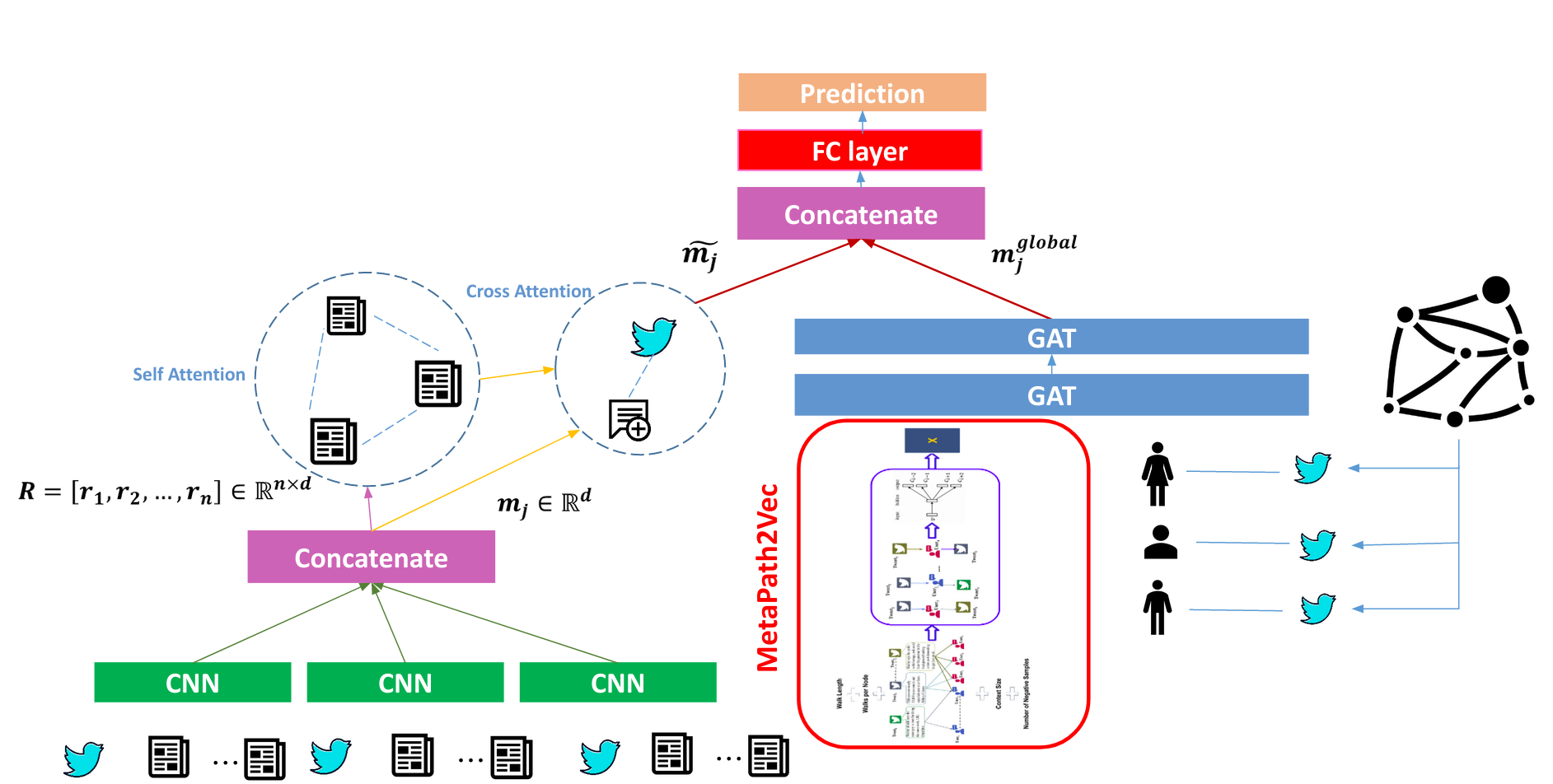}
  \end{center}
  \caption{The overall proposed architecture. The main difference between MGLAN compared to GLAN is that GLAN assigns node features by normal distribution when it does not have access to manually extracted features of Users and Tweets whereas MGLAN uses the output of MetaPath2Vec in each epoch as learned features of these node types.}
  \label{fig:proposed_method}
\end{figure*} 
\section{Related Work}\label{sec:related_work}
The previous models can be divided into three categories according to their approach:
\begin{itemize}
    \item \textbf{Content-based}: Extracting rich information from texts to learn specific writing styles that rumors inherently have is essential for their detection \cite{shu2019defend}. Text feature extraction can be done in supervised manners, i.e., TF-IDF and n-gram, or in unsupervised forms, i.e., embeddings from Word2Vec \cite{mikolov2013distributed}, LSTM \cite{hochreiter1997long}, GRU \cite{cho2014learning}, Transformers \cite{vaswani2017attention}, and BERT \cite{devlin2018bert}. These strategies can not be solely relied upon as text in social media usually has a short length; thus, they fail to capture the desired syntactic and semantic information needed to detect whether a piece of news is fake. 
    \item \textbf{User-based}: Manually scraped user features such as gender, age, nationality, and numbers of followers or followees are beneficial for rumor detection  \cite{shu2019role}.  \cite{castillo2011information} first used them to determine the credibility of information on Twitter. However, obtaining them is challenging as some social networks enforce restricting policies to access users' profiles or make them publicly available, like Twitter. Moreover, in terms of network representation, nodes are interconnected through edge interactions, and their feature vectors are not independent and identically distributed from one another \cite{nguyen2020fang}.  
    \item \textbf{Structure-based}: Leveraging the inherent structure of rumor propagation in social networks is another way that helps rumor detection. In  \cite{liu2018early}, the propagation path of news was modeled as multivariate time series. They can also be captured in a graph-setting environment considering the structural and semantic features of the news interaction network. In this setting, \textit{Graph Neural Networks(GNNs)} have shown a significant role in node-level, link-level, and graph-level prediction tasks. \cite{bian2020rumor} considered atop down and bottom-up approach to capture both propagation and dispersion of tweets using \textit{Graph Convolutional Networks(GCNs)} \cite{welling2016semi}. \cite{davoudi2022dss} proposed a deep hybrid model based on propagation and stance network of fake news and used \textit{node2vec} \cite{grover2016node2vec} for capturing structural propagation features on FakeNewsNet dataset \cite{shu2020fakenewsnet}. GLAN \cite{yuan2019jointly} offered a hybrid model using \textit{Graph Attention Networks(GATs)} \cite{velivckovic2017graph} to capture node-level representations of users and tweets.
\end{itemize}

Based on what elaborated above, GLAN offers a stable model to capture all three aspects but fails to assign initial features to nodes based on the difference in their types and their interconnectivity. When no initial features of users and tweets are accessible, It generates node features by the normal distribution in their implementation. This assumption can come from the fact that nodes and relations are independent, so initial feature generation using normal distribution would be sufficient. However, in reality, heterogeneous networks are scale-free in nature; in these networks, nodes and relations among them are not independent. This fact motivated us to modify a part of GLAN's architecture and use MetaPath2Vec \cite{dong2017metapath2vec} for extracting features for tweets and users in the propagation graph, discriminating among their node types. This modification has shown improvement in the performance of rumor detection when applied to three state-of-the-art datasets.  
\begin{figure*}[ht]
    \centering
    \setlength\abovecaptionskip{-0.1\baselineskip}
    \setlength\belowcaptionskip{-0.1\baselineskip}
    \includegraphics[width=0.8\textwidth]{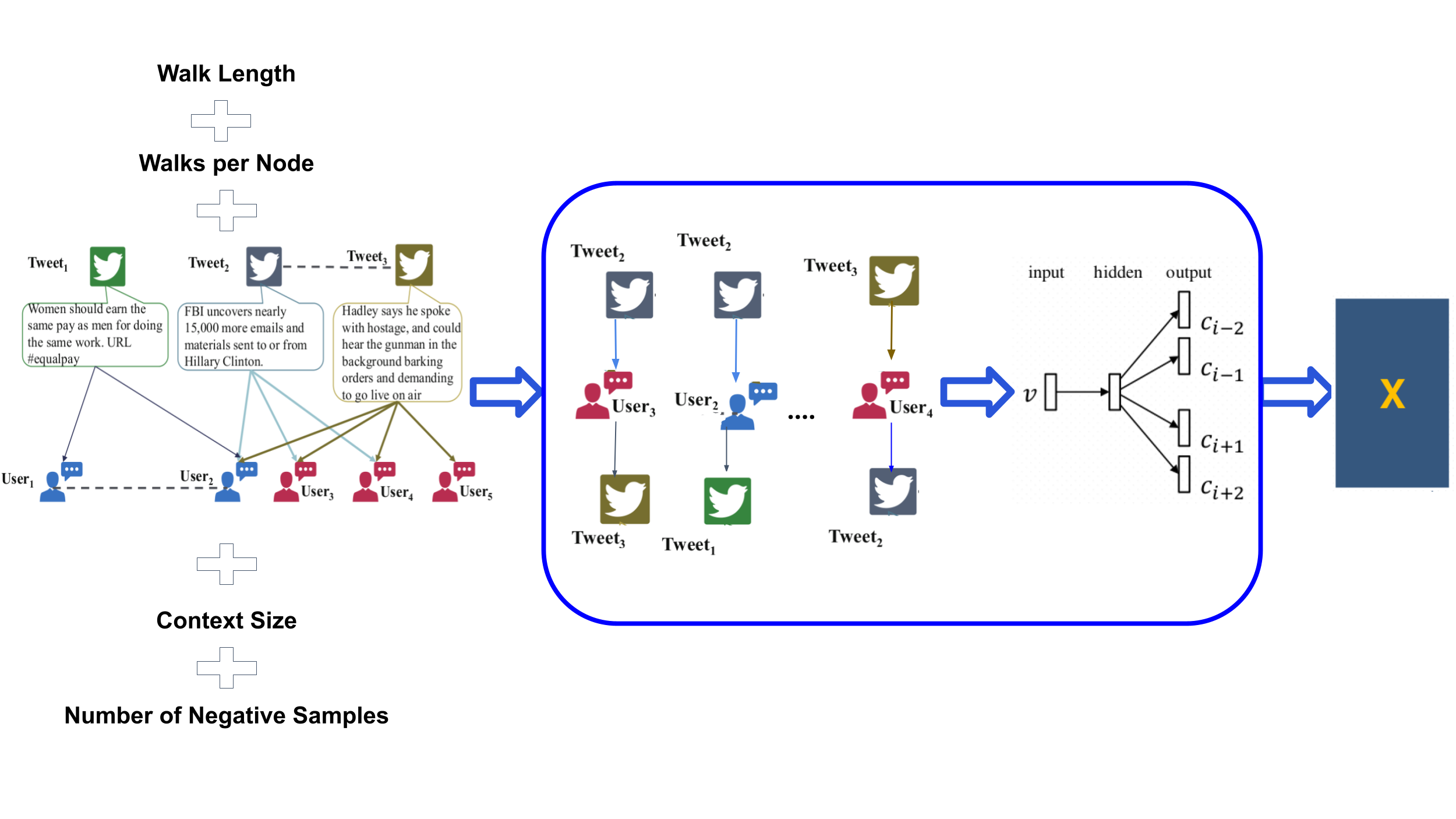}
  \caption{Architecture of MetaPath2Vec used as feature extraction module. The schema of heterogeneous graph is from \cite{yuan2019jointly} and \cite{dong2017metapath2vec}.}
  \label{fig:metapath2vec_architecture}
\end{figure*}
\section{Preliminaries and Problem Formulation}\label{sec:preliminaries}
IIn this section, we provide some basic definitions and then move on to formulate the rumor detection problem in this paper.

\begin{definition}
  \textit{Heterogeneous Network} \cite{sun2013mining}. A heterogeneous network is defined as a graph $G=(V, \mathcal{E})$ with a node type mapping function $\varphi: V\rightarrow \mathcal{A}$, and an edge type mapping function $\psi: \mathcal{E}\rightarrow \mathcal{R}$. $\mathcal{A}$ and $\mathcal{R}$ denote sets of node types and edge types, respectively, so that $\|\mathcal{A}\|+\|\mathcal{R}\| > 2$.
\end{definition}

\begin{definition}
	\textit{Network Schema} \cite{sun2011pathsim}. A network schema is a meta template for a heterogeneous network $G=(V, \mathcal{E})$, with a node type mapping function $\varphi: V\rightarrow \mathcal{A}$ and an edge type mapping function $\psi: \mathcal{E}\rightarrow \mathcal{R}$ defined over object types $\mathcal{A}$ denoted as $T_G=(\mathcal{A}, \mathcal{R})$.
\end{definition}

\begin{definition}
\textit{Meta Path} \cite{sun2011pathsim}. A meta path $P$ is a path on heterogeneous network with network schema $T_G=(\mathcal{A}, \mathcal{R})$ in the form of  
$A_1\xrightarrow{R_1}A_2\xrightarrow{R_2}...\xrightarrow{R_l}A_{l+1}$
 and $R$ is a compound relation $R=R_1\circ R_2\circ ... \circ R_l$ from $A_1$ to $A_l$.
\end{definition}

The formulation of our problem is quite similar to GLAN. Let $\mathcal{M}=\{m_1, m_2, ..., m_{|\mathcal{M}|}\}$ be the set of source news, and each source news has a total of $n$ retweets and replies denoted as $\mathcal{R}=\{r_1, r_2, ..., r_n\}$. We define the neighbors for a source news as $\mathcal{N}(m_i)=\{r_1, r_2,..., r_{\mathcal{N}_{(m_i)}}\}$. There are separate groups of global and local neighbors. We define replies of a source news as its local neighbors and its retweets as its global neighbors. The reason for assigning local and global terminologies is that replies of a tweet are independent from replies of other tweets, so they are categorized as local neighbors, but retweets diffuse through the whole network. Also, we define social media users as $\mathcal{U}=\{u_1, u_2, ..., u_{|\mathcal{U}|}\}$. Our objective is to learn a model $p(c=1|m_i, \mathcal{N}(m_i), \mathcal{U}; \theta)$ that takes a tweet and its neighbors as input. $c$ is the output and specifies the class to which the source news belongs. $\theta$ determines the model parameters.

\section{Proposed Method}\label{sec:proposed_method}
We aim to show that extracting meta path-based structural features from the heterogeneous network is conducive to rumor detection by having the propagation network without any initial knowledge of node-level features. 

In GLAN, initial node features for different node types were assigned by normal distribution when node-level features of tweets and users(\textit{e.g.} age, gender, number of likes, etc.) were unavailable. It is not the optimal way of feature extraction because it fails to take advantage of the rich information that the heterogeneous nature of the news propagation network provides. Recently, heterogeneous graph representation learning models like MetaPath2Vec\cite{dong2017metapath2vec} have shown promising success in extracting features using the concept of meta path in heterogeneous networks. Having this idea in mind, we decided to add a key module called \textit{Meta Path-based Feature Extraction} for better global feature extraction. Adding this module helps detect rumors more accurately in some evaluation metrics than GLAN. Fig.~\ref{fig:proposed_method} illustrates the whole architecture of MGLAN.

In the following, we describe our added component as well as GLAN modules in order to maintain integration throughout the paper. 
\subsection{Text Representation}

	Just like GLAN, we use word-level embeddings for word representation. $x_j\in\mathbb{R}^{d}$ is the $d$-dimensional embedding of $j$-$th$ word in text. We assume each text has fixed length $L$ represented as $x_{1:L}=[x_1; x_2; ...; x_L]$. Texts with more than $L$ words are truncated from the end till their length reaches $L$, and texts with lengths less than $L$ are zero-padded in the beginning till text length becomes $L$. Then, each text represented as $x_{1:L}$ is fed into three parallel CNNs \cite{zhang2015sensitivity} with $d/3$ dimension output to get semantic representation for each text. The size of the receptive field for each of the three CNNs is different with values $h\in\{3, 4, 5\}$. The $d/3$-dimensional outputs of each CNN are concatenated together and form the final $d$ dimensional representation. This process is performed on both source news text and replies of each source news separately, as demonstrated in Fig.~\ref{fig:proposed_method}. 

	\subsection{Local Relation Encoding}
	The term local relation of news refers to the relations that each news has with its surrounding neighbors in such a way that it is independent of the local relations of other news. In this section, we take the same approach as GLAN did. We use the attention mechanism to capture rich semantic relations of replies and source news and combine them into a single vector that encodes important aspects of source news and its replies. This procedure has two steps: 
	\begin{enumerate}
		\item{\textit{Self Attention}}: We use \textit{MultiHeadAttention} module as self attention with same inputs for parameters $Q$, $K$, and $V$\cite{vaswani2017attention}. If a piece of news has R replies, and each encoded in $d$-dimensional space, the output of the self-attention module is one $d$-dimensional embedding that has aggregated features of all the previous encoded replies denoted as $\widetilde{R}\in\mathbb{R}^{d}$.
		\begin{IEEEeqnarray}{cc}\label{eq:selfAttention}
	        \widetilde{R}=MultiHeadAttention(R, R, R)
	    \end{IEEEeqnarray}
		\item{\textit{Cross Attention}}: We apply cross attention to infuse source representation with the unified representation of its replies. The input $K$ of 
		\textit{MultiHeadAttention} is source news representation, $Q$ and $V$ are $\widetilde{R}$. This way, source representation can attend over its local neighbors to form a new $d$-dimensional local text representation denoted as $\widetilde{m_j}$ for news 
	\end{enumerate}

	\subsection{Meta Path-based Feature Extraction}
\begin{figure}[ht]
	    \centering
	    \setlength\abovecaptionskip{-0.1\baselineskip}
	    \setlength\belowcaptionskip{-0.1\baselineskip}
    	\includegraphics[width=0.28\textwidth]{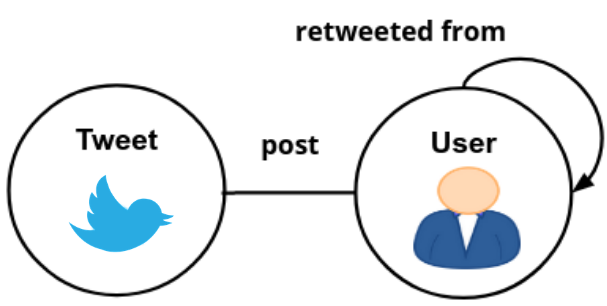}
    	\caption{\small News propagation network schema.}
    	\label{fig:network_schema}
    \end{figure}

\begin{figure}[ht]
       \centering
        \setlength\abovecaptionskip{-0.1\baselineskip}
        \setlength\belowcaptionskip{-0.1\baselineskip}
    	\includegraphics[width=0.25\textwidth]{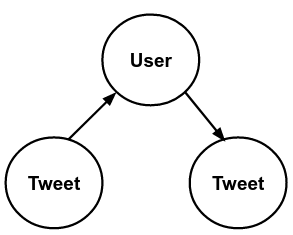}
    	\caption{\small Meta Path schema in news propagation network.}
    	\label{fig:metapath_schema}
    \end{figure}

	To capture efficient structural node representation while preserving inherent discrimination among each node type, we modeled the news propagation network as a heterogeneous graph with two node types: User and Tweet. Fig.~\ref{fig:network_schema} shows the network schema of news propagation on Twitter. At first, a Tweet is published by a user, and other users participate in retweeting it from publishing users or other retweeters. 
	To simplify the schema, we decided to behave post and retweet relations as one relation called \textit{spread} that contains both retweet and post relations. It means a user participates in \textit{spreading} tweets and tweets \textit{are spread by} users in the network. Finally, we define the meta path schema for the rumor detection problem in  Fig.~\ref{fig:metapath_schema}. We feed meta path relations schema alongside the network edges into MetaPath2Vec architecture as illustrated in Fig.~\ref{fig:metapath2vec_architecture}. MetaPath2Vec is robust among heterogeneous representation learning methods to discern structural and semantic correlation among different node types. In order to pay more attention to early diffuser users,  weights between edges that connect users and tweets were assigned as follows: 
	\begin{IEEEeqnarray}{cc}\label{edge_weights}
	    \mathnormal{w}(u_i, m_j) = \frac{1}{\mathnormal{max}(0, t) +1}
    \end{IEEEeqnarray}
    
    $t$ is time elapsed(in minutes) after a tweet was published \cite{yuan2019jointly}. This weighting helps us to modify the random walk process of MetaPath2Vec. In the original paper, a node in the next random walk step was selected by the uniform distribution.
    In our work, a node is selected by weighted distribution according to the following:
    \begin{equation}\label{eq:WeightedRandomWalk}
    \begin{aligned}
    p(v^{i+1} | v^i, P) = 
    \begin{cases}
    \frac{w_{v_iv_{i+1}}}{\Sigma w_{v_i u}} & \mathnormal{(v^{i+1}, v^i_t)\in E,\phi(v^{i+1})=t+1}\\
    0 & \mathnormal{(v^{i+1}, v^i_t)\in E,\phi(v^{i+1})\neq t+1} \\
    0 & \mathnormal{(v^{i+1}, v^i_t)\notin E}\end{cases}
    \end{aligned}
    \end{equation}
where $u$ is selected from all neighbors of $v_i$, $u\in\mathcal{N}(v_i)$. The output of MetaPath2Vec is $d$-dimensional node-level representation of all nodes $\mathcal{X}\in\mathbb{R}^{n\times d}$. 
	
\subsection{Global Relation Encoding}
	Just like GLAN, we used two GATs to capture additional meaningful representations of different nodes. The output of MetaPath2Vec $\mathcal{X}$ is the input of the first GAT. In the first one, \textit{MultiHeadAttention} with k=8 heads are enabled to stabilize the training process. In the second one, \textit{MultiHeadAttention} is disabled. The output of the global relation encoding module is $m_j^{global}$. 
\subsection{Rumor Classification}
	In the classification module, we concatenate $\widetilde{m_j}$ and $m_j^{global}$ and pass it through a fully-connected linear layer. Then by applying \textit{softmax} and choosing the maximum probability, we can classify each source news:
	\begin{equation}\label{eq:GLANLikelihood}
	p_i(c|m_i, \mathcal{N}_{m_i}, \mathcal{U};\theta)=\textbf{softmax}(\textbf{W}[\widetilde{m_j}, m^{global}_j]+b)
	\end{equation}
	$\textbf{W}\in\mathbb{R}^{2d\times |c|}$ is weight parameter of linear layer and $b$ is bias.
	Cross entropy loss is used to classify each piece of news:
	\begin{equation}
	J(c^{(i)}| D, \mathcal{U}_i; \theta) = - \sum_{i}y_i \log p_i(c|m_i, \mathcal{N}_{m_i},\mathcal{U};\theta)
	\end{equation}
	$y_i$ is the probability of source news belonging to class $i$.

\section{Evaluation}\label{sec:evaluation}
\begin{figure*}[t]
    \centering
    \subfloat[Twitter15]{\includegraphics[width=0.34\linewidth]{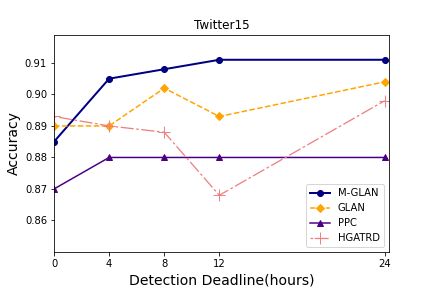}
    \label{fig:twitter15_results}}
    \subfloat[Twitter16]{\includegraphics[width=0.34\linewidth]{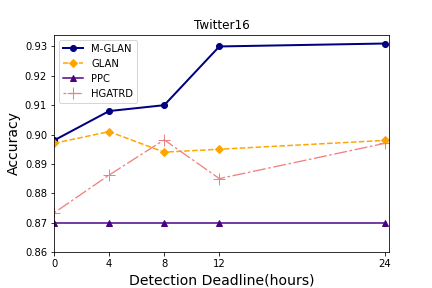}
    \label{fig:twitter16_results}}
    \subfloat[Weibo]{\includegraphics[width=0.34\linewidth]{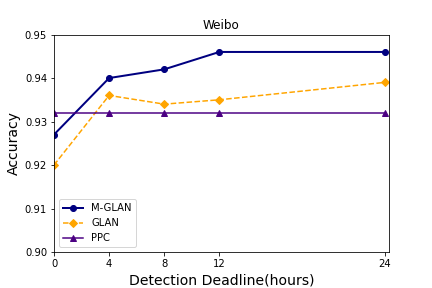}
    \label{fig:weibo_results}}
    \caption{Time limit vs Accuracy in Twitter15, Twitter16 and Weibo.}
    \label{fig:early_detection_results}
\end{figure*}
In this section, we perform experiments on three state of the art datasets for rumor detection. We show that our proposed architecture outperforms similar state-of-the-art models.
\subsection{Datasets}
We analyzed MGLAN on three well-known datasets: Twitter15\cite{ma2017detect}, Twitter16\cite{ma2017detect}, and Weibo\cite{ma2016detecting}. Data are scraped from Twitter and Weibo social networks, respectively. Twiter15 and Twitter16 consist of four classes: "NR"(non-rumor), "FR"(false-rumor), "UR"(unverified-rumor), and "TR"(true-rumor). The difference between "FR" and "TR" is that a true label is assigned to a source tweet if it expresses a denial type of stance; otherwise, it is labeled as false \cite{ma2017detect}. The Weibo dataset has only binary labels "NR" and "FR." Table \ref{tab:dataset_statistics} provides a statistical perspective of datasets.

\begin{table}[ht]
    \begin{center}
    \caption{Dataset Statistics}
    \label{tab:dataset_statistics}
\begin{tabular}{c c c c} 
     \hline
     \textbf{Statistics} & \textbf{Twitter15} & \textbf{Twitter16} & \textbf{Weibo} \\ 
     \hline\hline
     \# tweets & \numprint{1490} & 818 & \numprint{4664} \\ 
     \hline
     \# users & \numprint{276663} & \numprint{173487} & \numprint{2746818} \\
     \hline
     \# posts & \numprint{331612} & \numprint{204820} & \numprint{3805656} \\
     \hline
     \# NR & 374 & 205 & \numprint{2351} \\ 
     \hline
     \# FR & 370 & 205 & \numprint{2313} \\ 
     \hline
     \# UR & 374 & 203 & 0 \\
     \hline
     \# TR & 372 & 203 & 0 \\
     \hline
    \end{tabular}
    \end{center}
\end{table}

\subsection{Baseline Models}
In this section, we introduce previously proposed architectures for the task of rumor detection and compare their performance to MGLAN.
\begin{itemize}
    \item \textbf{GLAN}: Our baseline. It encodes both local and global relations of the heterogeneous network.
    \item \textbf{HGATRD} \cite{huang2020heterogeneous}: It models the propagation network as a heterogeneous one with tweet, user, and word as node types. It decomposes the graph into tweet-word and tweet-user subgraphs and performs attention mechanisms for each subgraph.
    \item \textbf{SMAN} \cite{yuan2020early}: Based on the news that each user participates in spreading, it assigns a credibility score to each user and uses it as weakly supervised information. It then uses MultiHeadAttention to learn to classify each source news.
    \item \textbf{GCAN} \cite{lu2020gcan}: It creates user communication graphs for all tweets, and GCNs compute their graph-level embeddings. On the other hand, It uses CNNs to model the sequential retweet path of each tweet. Finally, It concatenates respective outputs after applying attention mechanisms and passes them to the classifier.
\item \textbf{PPC} \cite{liu2018early}: It is based on modeling the propagation path of tweets as a multivariate time series, then builds a propagation path classifier using both CNNs and RNNs.
\end{itemize}
\subsection{Parameter Settings}
MGLAN is an extension to GLAN implemented with Pytorch \cite{paszke2017automatic}, and Pytorch Geometric \cite{fey2019fast}. For true comparison, we did not change any of GLAN's hyperparameters. In MetaPath2Vec, we set walk length to 100, context size to 7, number of walks per node to 5, and number of negative samples to 3. The output of MetaPath2Vec has 256 dimensions. The dimension of the first GAT's input and output is 256 and 300. The second GAT, both input and output  have 300 dimensions.
\subsection{Results and Analysis}
\begin{figure*}[t]
    \centering
    \subfloat[GLAN embeddings in Twitter15]{\includegraphics[width=0.34\linewidth]{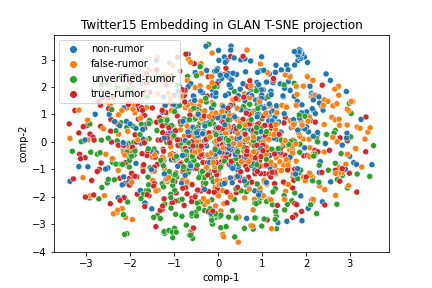}
    \label{fig:GLAN_twitter15_results}}
    \subfloat[GLAN embeddings in Twitter16]{\includegraphics[width=0.34\linewidth]{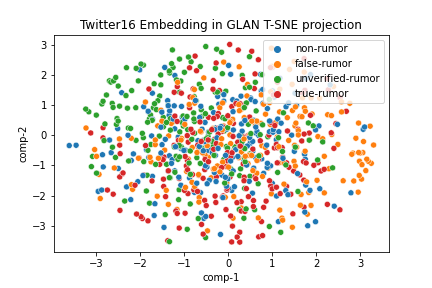}
    \label{fig:GLAN_twitter16_results}}
    \subfloat[GLAN embeddings in Weibo]{\includegraphics[width=0.34\linewidth]{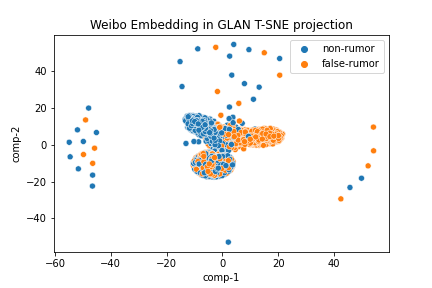}
    \label{fig:GLAN_weibo_results}}
    
    \subfloat[MGLAN embeddings in Twitter15]{\includegraphics[width=0.34\linewidth]{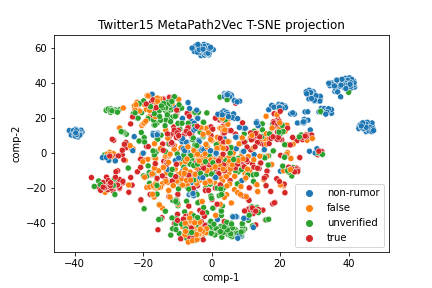}
    \label{fig:MGLAN_twitter15_results}}
    \subfloat[MGLAN embeddings in Twitter16]{\includegraphics[width=0.34\linewidth]{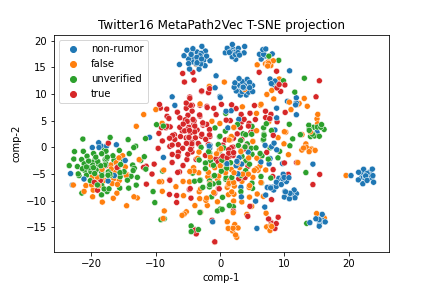}
    \label{fig:MGLAN_twitter16_results}}
    \subfloat[MGLAN embeddings in Weibo]{\includegraphics[width=0.34\linewidth]{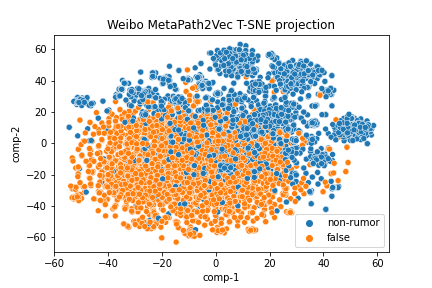}
    \label{fig:MGLAN_weibo_results}}
    
    \caption{Representations of tweet nodes of the heterogeneous news network in MGLAN vs. GLAN. Utilizing MetaPath2Vec could capture meaningful structural features since nodes belonging to the same group almost form a cluster, whereas in GLAN, in cases where a single normal distribution assigned node features, clusters are not formed.}
    \label{fig:t-SNE_projection}
\end{figure*}

This section shows that MGLAN is robust enough to detect rumors better than GLAN and other models. Tables \ref{tab:Twitter15Results}, \ref{tab:Twitter16Results}, and \ref{tab:WeiboResults} show that MGLAN performs better in almost all classes compared to other models by accuray and F1 score metric; meaning extracting the heterogeneous-based representations play an essential role for rumor detection. 

However, in Twitter15 in UR class and Twitter16 in FR class, MGLAN failed to outperform SMAN due to computing a credibility score for users from the existing dataset and using it as weakly supervised information. MGLAN is not reliant on hand-crafted features and, therefore, can be applied in inductive settings where new accounts are created and participate in rumor dissemination. Also, we saw the slightest improvement in accuracy and F1 score in the Weibo dataset. It is related to the inherent characteristic of the dataset; source texts and their corresponding replies are distinguishable enough that considering a meta path-based heterogeneous feature extraction performs slightly better in this case. 

In conclusion, Incorporating MetaPath2Vec and capturing heterogeneous structural features in news propagation networks shows improvement over previous models. We will discuss this assumption from another perspective in section
\ref{metapath2vec_robustness}.

    \begin{table}[ht]
	\caption{Twitter15 Results}
	\label{tab:Twitter15Results}
	\centering
	\begin{tabular}{|c|c|c|c|c|c|}
		\hline Model & Accuracy & NR & FR & UR & TR\\ 
		& & F1 & F1 & F1 & F1\\
		\hline PPC & 0.842 & 0.811 & 0.875& 0.79 & 0.818\\ 
		\hline GLAN & 0.900 & 0.926  & 0.909 & 0.834 & 0.918\\ 
		\hline SMAN & 0.895 & 0.896  & 0.888& \textbf{0.921} & 0.867\\ 
		\hline HGATRD & 0.892 & 0.918 & 0.887 & 0.909 &0.84 \\ 
		\hline \textbf{MGLAN} & \textbf{0.917} & \textbf{0.943} & \textbf{0.928}& 0.874 & \textbf{0.921} \\ 
		\hline 
	\end{tabular} 
\end{table}
\begin{table}[ht]
	\caption{Twitter16 Results}
	\label{tab:Twitter16Results}
	\centering
	\begin{tabular}{|c|c|c|c|c|c|}
		\hline Model& Accuracy &NR&FR&UR& TR\\ 
		& & F1 & F1 & F1 & F1\\
		\hline PPC &0.863 &0.843 &0.898&0.837&0.820 \\ 
		\hline GLAN & 0.902 & 0.921  & 0.869 & 0.878 & 0.968\\ 
		\hline SMAN & 0.911 & 0.929 & \textbf{0.952}& 0.931 & 0.857\\ 
		\hline HGATRD & 0.886 & 0.911 & 0.853& 0.847 & 0.927 \\ 
		\hline \textbf{MGLAN} & \textbf{0.936} & \textbf{0.938} & 0.932 & \textbf{0.935} & \textbf{0.94} \\ 
		\hline 
	\end{tabular} 
\end{table}
\begin{table}[ht]
	\caption{Weibo Resulsts}
	\label{tab:WeiboResults}
	\centering
	\begin{tabular}{|c|c|c|c|}
		\hline Model & Accuracy & NR & FR \\
		& & F1 & F1 \\ 
		\hline PPC & 0.921 & 0.918 & 0.923\\ 
		\hline GLAN & 0.942& 0.941  & 0.943\\ 
		\hline SMAN & 0.937 & 0.945 & 0.946 \\ 
		\hline \textbf{MGLAN} & \textbf{0.952} &  \textbf{0.951} &\textbf{0.952}\\
		\hline 
	\end{tabular} 
\end{table}

\subsection{Early Rumor Detection}
To prove MGLAN can detect rumors in their early phases of creation, we decided to 
restrict the time after a tweet is published and examine how MGLAN performs against other models. As shown in Fig.~\ref{fig:twitter15_results}, Fig.~\ref{fig:twitter16_results}, and Fig.~\ref{fig:weibo_results}. The horizontal axis shows elapsed time since a tweet was published, and the vertical axis represents the accuracy of models. It is apparent in all three figures that right after the tweet was published, MGLAN and GLAN had the same accuracy because the global news propagation network is not shaped. After a short time during which the structure of the propagation network is formed, MGLAN outperforms GLAN and other models.

\subsection{Robustness of MGLAN}
\label{metapath2vec_robustness}
To elaborate on the effectiveness of using MetaPath2Vec from a different perspective, we decided to plot the t-SNE\cite{van2008visualizing} projections of heterogeneous embeddings in MGLAN and GLAN, respectively. As shown in Fig.~\ref{fig:t-SNE_projection}, in MGLAN, projected MetaPath2Vec embeddings in all datasets can cluster tweets of each class to a reasonable extent. For example, if two tweets belong to one class, MetaPath2Vec has similar $d$-dimensional embeddings for them. In contrast, in GLAN, by assigning node features using the normal distribution when user or tweet metadata is not accessible, projections of each class are not distinguishable from one another, especially in Twitter15 and Twitter16.
\section{Conclusions and Future Work}\label{sec:conclusion}
In this work, we proposed improving GLAN architecture using meta path-based features in heterogeneous news propagation networks called MGLAN. For implementation, we used MetaPath2Vec to extract structural features from the heterogeneous graph and used its output embeddings to GLAN's global relation encoding module. Also, experiments on three state-of-the-art datasets proved the strength of MGLAN.
For future work, we can extend our proposed idea in several cases: 
\begin{itemize}
    \item We can scrape a new dataset from Twitter while gathering metadata for different node types and therefore define new relations such as follower-followee and mentions. It allows us to incorporate a variety of meta paths and determine which one plays a more critical role in extracting richer information from the heterogeneous network.
 
    \item MetaPath2Vec can focus on only one meta path at one time. Newer methods like MAGNN \cite{fu2020magnn} can simultaneously consider different meta paths and choose the best one for each heterogeneous graph.
\end{itemize}
\section*{Acknowledgment}
Mostafa Salehi was supported by a grant from IPM, Iran (No. CS1401-4-162).
\bibliographystyle{IEEEtran}

\bibliography{IEEErefs}

\end{document}